\documentclass[aps,prl,reprint,superscriptaddress,showpacs,nopreprintnumbers,
               nofootinbib,longbibliography]{revtex4-1}

\usepackage{amsmath}
\usepackage{graphicx, color}
\usepackage{hyperref}
\usepackage{etoolbox}
\usepackage[modulo]{lineno}

\definecolor{redHL}{rgb}{1.0,0.9,0.9}
\newcommand{\fixme}[1]{{#1}}

\newtoggle{endauthorlist}
\toggletrue{endauthorlist}

\begin{document}

\title{GW150914: The Advanced LIGO Detectors in the Era of First Discoveries}

\iftoggle{endauthorlist}{
  \let\mymaketitle\maketitle
  \let\myauthor\author
  \let\myaffiliation\affiliation
  \author{B. P. Abbott {\it et al.}}
  \thanks{Full author list given at the end of the article.}
  \collaboration{LIGO Scientific Collaboration and Virgo Collaboration}
  \noaffiliation
}{
  
\author{%
B.~P.~Abbott,$^{1}$  
R.~Abbott,$^{1}$  
T.~D.~Abbott,$^{2}$  
M.~R.~Abernathy,$^{1}$  
F.~Acernese,$^{3,4}$
K.~Ackley,$^{5}$  
C.~Adams,$^{6}$  
T.~Adams,$^{7}$
P.~Addesso,$^{3}$  
R.~X.~Adhikari,$^{1}$
% ...
\\
\medskip
(LIGO Scientific Collaboration and Virgo Collaboration)
\\
\medskip
{{}$^{\dag}$Deceased, May 2015. {}$^{\ddag}$Deceased, March 2015. }%
}\noaffiliation
\affiliation {LIGO, California Institute of Technology, Pasadena, CA 91125, USA }
\affiliation {Louisiana State University, Baton Rouge, LA 70803, USA }
\affiliation {Universit\`a di Salerno, Fisciano, I-84084 Salerno, Italy }
\affiliation {INFN, Sezione di Napoli, Complesso Universitario di Monte S.Angelo, I-80126 Napoli, Italy }
\affiliation {University of Florida, Gainesville, FL 32611, USA }
\affiliation {LIGO Livingston Observatory, Livingston, LA 70754, USA }
\affiliation {Laboratoire d'Annecy-le-Vieux de Physique des Particules (LAPP), Universit\'e Savoie Mont Blanc, CNRS/IN2P3, F-74941 Annecy-le-Vieux, France }
\affiliation {Albert-Einstein-Institut, Max-Planck-Institut f\"ur Gravi\-ta\-tions\-physik, D-30167 Hannover, Germany }
\affiliation {Nikhef, Science Park, 1098 XG Amsterdam, Netherlands }
\affiliation {LIGO, Massachusetts Institute of Technology, Cambridge, MA 02139, USA }
\affiliation {Instituto Nacional de Pesquisas Espaciais, 12227-010 S\~{a}o Jos\'{e} dos Campos, S\~{a}o Paulo, Brazil }
\affiliation {INFN, Gran Sasso Science Institute, I-67100 L'Aquila, Italy }
\affiliation {INFN, Sezione di Roma Tor Vergata, I-00133 Roma, Italy }
\affiliation {Inter-University Centre for Astronomy and Astrophysics, Pune 411007, India }
\affiliation {International Centre for Theoretical Sciences, Tata Institute of Fundamental Research, Bangalore 560012, India }
\affiliation {University of Wisconsin-Milwaukee, Milwaukee, WI 53201, USA }
\affiliation {Leibniz Universit\"at Hannover, D-30167 Hannover, Germany }
\affiliation {Universit\`a di Pisa, I-56127 Pisa, Italy }
\affiliation {INFN, Sezione di Pisa, I-56127 Pisa, Italy }
\affiliation {Australian National University, Canberra, Australian Capital Territory 0200, Australia }
\affiliation {The University of Mississippi, University, MS 38677, USA }
\affiliation {California State University Fullerton, Fullerton, CA 92831, USA }
\affiliation {LAL, Universit\'e Paris-Sud, CNRS/IN2P3, Universit\'e Paris-Saclay, 91400 Orsay, France }
\affiliation {Chennai Mathematical Institute, Chennai 603103, India }
\affiliation {Universit\`a di Roma Tor Vergata, I-00133 Roma, Italy }
\affiliation {University of Southampton, Southampton SO17 1BJ, United Kingdom }
\affiliation {Universit\"at Hamburg, D-22761 Hamburg, Germany }
\affiliation {INFN, Sezione di Roma, I-00185 Roma, Italy }
\affiliation {Albert-Einstein-Institut, Max-Planck-Institut f\"ur Gravitations\-physik, D-14476 Potsdam-Golm, Germany }
\affiliation {APC, AstroParticule et Cosmologie, Universit\'e Paris Diderot, CNRS/IN2P3, CEA/Irfu, Observatoire de Paris, Sorbonne Paris Cit\'e, F-75205 Paris Cedex 13, France }
\affiliation {Montana State University, Bozeman, MT 59717, USA }
\affiliation {Universit\`a di Perugia, I-06123 Perugia, Italy }
\affiliation {INFN, Sezione di Perugia, I-06123 Perugia, Italy }
\affiliation {European Gravitational Observatory (EGO), I-56021 Cascina, Pisa, Italy }
\affiliation {Syracuse University, Syracuse, NY 13244, USA }
\affiliation {SUPA, University of Glasgow, Glasgow G12 8QQ, United Kingdom }
\affiliation {LIGO Hanford Observatory, Richland, WA 99352, USA }
\affiliation {Wigner RCP, RMKI, H-1121 Budapest, Konkoly Thege Mikl\'os \'ut 29-33, Hungary }
\affiliation {Columbia University, New York, NY 10027, USA }
\affiliation {Stanford University, Stanford, CA 94305, USA }
\affiliation {Universit\`a di Padova, Dipartimento di Fisica e Astronomia, I-35131 Padova, Italy }
\affiliation {INFN, Sezione di Padova, I-35131 Padova, Italy }
\affiliation {CAMK-PAN, 00-716 Warsaw, Poland }
\affiliation {Astronomical Observatory Warsaw University, 00-478 Warsaw, Poland }
\affiliation {University of Birmingham, Birmingham B15 2TT, United Kingdom }
\affiliation {Universit\`a degli Studi di Genova, I-16146 Genova, Italy }
\affiliation {INFN, Sezione di Genova, I-16146 Genova, Italy }
\affiliation {RRCAT, Indore MP 452013, India }
\affiliation {Faculty of Physics, Lomonosov Moscow State University, Moscow 119991, Russia }
\affiliation {SUPA, University of the West of Scotland, Paisley PA1 2BE, United Kingdom }
\affiliation {University of Western Australia, Crawley, Western Australia 6009, Australia }
\affiliation {Department of Astrophysics/IMAPP, Radboud University Nijmegen, P.O. Box 9010, 6500 GL Nijmegen, Netherlands }
\affiliation {Artemis, Universit\'e C\^ote d'Azur, CNRS, Observatoire C\^ote d'Azur, CS 34229, Nice cedex 4, France }
\affiliation {MTA E\"otv\"os University, ``Lendulet'' Astrophysics Research Group, Budapest 1117, Hungary }
\affiliation {Institut de Physique de Rennes, CNRS, Universit\'e de Rennes 1, F-35042 Rennes, France }
\affiliation {Washington State University, Pullman, WA 99164, USA }
\affiliation {Universit\`a degli Studi di Urbino ``Carlo Bo,'' I-61029 Urbino, Italy }
\affiliation {INFN, Sezione di Firenze, I-50019 Sesto Fiorentino, Firenze, Italy }
\affiliation {University of Oregon, Eugene, OR 97403, USA }
\affiliation {Laboratoire Kastler Brossel, UPMC-Sorbonne Universit\'es, CNRS, ENS-PSL Research University, Coll\`ege de France, F-75005 Paris, France }
\affiliation {VU University Amsterdam, 1081 HV Amsterdam, Netherlands }
\affiliation {University of Maryland, College Park, MD 20742, USA }
\affiliation {Center for Relativistic Astrophysics and School of Physics, Georgia Institute of Technology, Atlanta, GA 30332, USA }
\affiliation {Institut Lumi\`{e}re Mati\`{e}re, Universit\'{e} de Lyon, Universit\'{e} Claude Bernard Lyon 1, UMR CNRS 5306, 69622 Villeurbanne, France }
\affiliation {Laboratoire des Mat\'eriaux Avanc\'es (LMA), IN2P3/CNRS, Universit\'e de Lyon, F-69622 Villeurbanne, Lyon, France }
\affiliation {Universitat de les Illes Balears, IAC3---IEEC, E-07122 Palma de Mallorca, Spain }
\affiliation {Universit\`a di Napoli ``Federico II,'' Complesso Universitario di Monte S.Angelo, I-80126 Napoli, Italy }
\affiliation {NASA/Goddard Space Flight Center, Greenbelt, MD 20771, USA }
\affiliation {Canadian Institute for Theoretical Astrophysics, University of Toronto, Toronto, Ontario M5S 3H8, Canada }
\affiliation {Tsinghua University, Beijing 100084, China }
\affiliation {Texas Tech University, Lubbock, TX 79409, USA }
\affiliation {The Pennsylvania State University, University Park, PA 16802, USA }
\affiliation {National Tsing Hua University, Hsinchu City, 30013 Taiwan, Republic of China }
\affiliation {Charles Sturt University, Wagga Wagga, New South Wales 2678, Australia }
\affiliation {University of Chicago, Chicago, IL 60637, USA }
\affiliation {Caltech CaRT, Pasadena, CA 91125, USA }
\affiliation {Korea Institute of Science and Technology Information, Daejeon 305-806, Korea }
\affiliation {Carleton College, Northfield, MN 55057, USA }
\affiliation {Universit\`a di Roma ``La Sapienza,'' I-00185 Roma, Italy }
\affiliation {University of Brussels, Brussels 1050, Belgium }
\affiliation {Sonoma State University, Rohnert Park, CA 94928, USA }
\affiliation {Northwestern University, Evanston, IL 60208, USA }
\affiliation {University of Minnesota, Minneapolis, MN 55455, USA }
\affiliation {The University of Melbourne, Parkville, Victoria 3010, Australia }
\affiliation {The University of Texas Rio Grande Valley, Brownsville, TX 78520, USA }
\affiliation {The University of Sheffield, Sheffield S10 2TN, United Kingdom }
\affiliation {University of Sannio at Benevento, I-82100 Benevento, Italy and INFN, Sezione di Napoli, I-80100 Napoli, Italy }
\affiliation {Montclair State University, Montclair, NJ 07043, USA }
\affiliation {Universit\`a di Trento, Dipartimento di Fisica, I-38123 Povo, Trento, Italy }
\affiliation {INFN, Trento Institute for Fundamental Physics and Applications, I-38123 Povo, Trento, Italy }
\affiliation {Cardiff University, Cardiff CF24 3AA, United Kingdom }
\affiliation {National Astronomical Observatory of Japan, 2-21-1 Osawa, Mitaka, Tokyo 181-8588, Japan }
\affiliation {School of Mathematics, University of Edinburgh, Edinburgh EH9 3FD, United Kingdom }
\affiliation {Indian Institute of Technology, Gandhinagar Ahmedabad Gujarat 382424, India }
\affiliation {Institute for Plasma Research, Bhat, Gandhinagar 382428, India }
\affiliation {University of Szeged, D\'om t\'er 9, Szeged 6720, Hungary }
\affiliation {Embry-Riddle Aeronautical University, Prescott, AZ 86301, USA }
\affiliation {University of Michigan, Ann Arbor, MI 48109, USA }
\affiliation {Tata Institute of Fundamental Research, Mumbai 400005, India }
\affiliation {American University, Washington, D.C. 20016, USA }
\affiliation {University of Massachusetts-Amherst, Amherst, MA 01003, USA }
\affiliation {University of Adelaide, Adelaide, South Australia 5005, Australia }
\affiliation {West Virginia University, Morgantown, WV 26506, USA }
\affiliation {University of Bia{\l }ystok, 15-424 Bia{\l }ystok, Poland }
\affiliation {SUPA, University of Strathclyde, Glasgow G1 1XQ, United Kingdom }
\affiliation {IISER-TVM, CET Campus, Trivandrum Kerala 695016, India }
\affiliation {Institute of Applied Physics, Nizhny Novgorod, 603950, Russia }
\affiliation {Pusan National University, Busan 609-735, Korea }
\affiliation {Hanyang University, Seoul 133-791, Korea }
\affiliation {NCBJ, 05-400 \'Swierk-Otwock, Poland }
\affiliation {IM-PAN, 00-956 Warsaw, Poland }
\affiliation {Rochester Institute of Technology, Rochester, NY 14623, USA }
\affiliation {Monash University, Victoria 3800, Australia }
\affiliation {Seoul National University, Seoul 151-742, Korea }
\affiliation {University of Alabama in Huntsville, Huntsville, AL 35899, USA }
\affiliation {ESPCI, CNRS, F-75005 Paris, France }
\affiliation {Universit\`a di Camerino, Dipartimento di Fisica, I-62032 Camerino, Italy }
\affiliation {Southern University and A\&M College, Baton Rouge, LA 70813, USA }
\affiliation {College of William and Mary, Williamsburg, VA 23187, USA }
\affiliation {Instituto de F\'\i sica Te\'orica, University Estadual Paulista/ICTP South American Institute for Fundamental Research, S\~ao Paulo SP 01140-070, Brazil }
\affiliation {University of Cambridge, Cambridge CB2 1TN, United Kingdom }
\affiliation {IISER-Kolkata, Mohanpur, West Bengal 741252, India }
\affiliation {Rutherford Appleton Laboratory, HSIC, Chilton, Didcot, Oxon OX11 0QX, United Kingdom }
\affiliation {Whitman College, 345 Boyer Avenue, Walla Walla, WA 99362 USA }
\affiliation {National Institute for Mathematical Sciences, Daejeon 305-390, Korea }
\affiliation {Hobart and William Smith Colleges, Geneva, NY 14456, USA }
\affiliation {Janusz Gil Institute of Astronomy, University of Zielona G\'ora, 65-265 Zielona G\'ora, Poland }
\affiliation {Andrews University, Berrien Springs, MI 49104, USA }
\affiliation {Universit\`a di Siena, I-53100 Siena, Italy }
\affiliation {Trinity University, San Antonio, TX 78212, USA }
\affiliation {University of Washington, Seattle, WA 98195, USA }
\affiliation {Kenyon College, Gambier, OH 43022, USA }
\affiliation {Abilene Christian University, Abilene, TX 79699, USA }

}

\date[\relax]{\today}

\begin{abstract}
Following a major upgrade, the two advanced detectors of the Laser Interferometer Gravitational-wave Observatory (LIGO) held their first observation run between September 2015 and January 2016. With a strain sensitivity of $10^{-23}/\sqrt{\mathrm{Hz}}$ at 100\,Hz, the product of observable volume and measurement time exceeded that of all previous runs within the first \fixme{16} days of coincident observation. On September 14th, 2015 the Advanced LIGO detectors observed a transient gravitational-wave signal determined to be the coalescence of two black holes~\cite{detection:2016}, launching the era of gravitational-wave astronomy. The event, GW150914, was observed with a combined signal-to-noise ratio of \fixme{24} in coincidence by the two detectors. Here we present the main features of the detectors that enabled this observation. At full sensitivity, the Advanced LIGO detectors are designed to deliver another factor of \fixme{three} improvement in the signal-to-noise ratio for binary black hole systems similar in masses to GW150914.
\end{abstract}


\maketitle

{\it Introduction} --- On September 14th, 2015, both Advanced LIGO detectors in the USA, H1 in Hanford, Washington and L1 in Livingston, Lousiana, made the first direct measurement of gravitational waves~\cite{detection:2016}. The event, GW150914, was determined to be the merger of two black holes, with masses of  \fixme{36}\,$M_\odot$ and \fixme{29}\,$M_\odot$, into a black hole of approximately \fixme{62}\,$M_\odot$~\cite{compParamEst:2016}. \fixme{3.0} solar masses of energy (\fixme{$\simeq\!5.4\times\! 10^{47}$\,J}) was radiated in gravitational waves. The gravitational waves from this event, which occurred at a distance of \fixme{$\simeq\!410\, {\rm Mpc} \simeq 1.3 \times\! 10^9 \,{\rm light~years}$}, changed the separation between the test masses by $\simeq\!4\times\! 10^{-18}$\,m, about one 200th of a proton radius.

The Advanced LIGO detectors, multi-kilometer Michelson-based interferometers~\cite{aligo}, came online in September 2015, after a major upgrade targeting a factor of 10 sensitivity improvement over initial detectors~\cite{iligo,ivirgo}. While not yet at design sensitivity during their first observation run, they have already exceeded the strain sensitivity of the initial detectors across the entire frequency band, significantly surpassing the past discovery potential~\cite{compAstro:2016,compGR:2016, compStoch:2016,compLoc:2016, compNeutrino:2016}. This paper describes the Advanced LIGO detectors, as well as their current and final design sensitivity, at the inception of gravitational-wave astronomy.

\begin{figure*}[t]
\begin{center}
    \begin{minipage}{0.495\textwidth}
        \includegraphics[width = 0.95\textwidth]{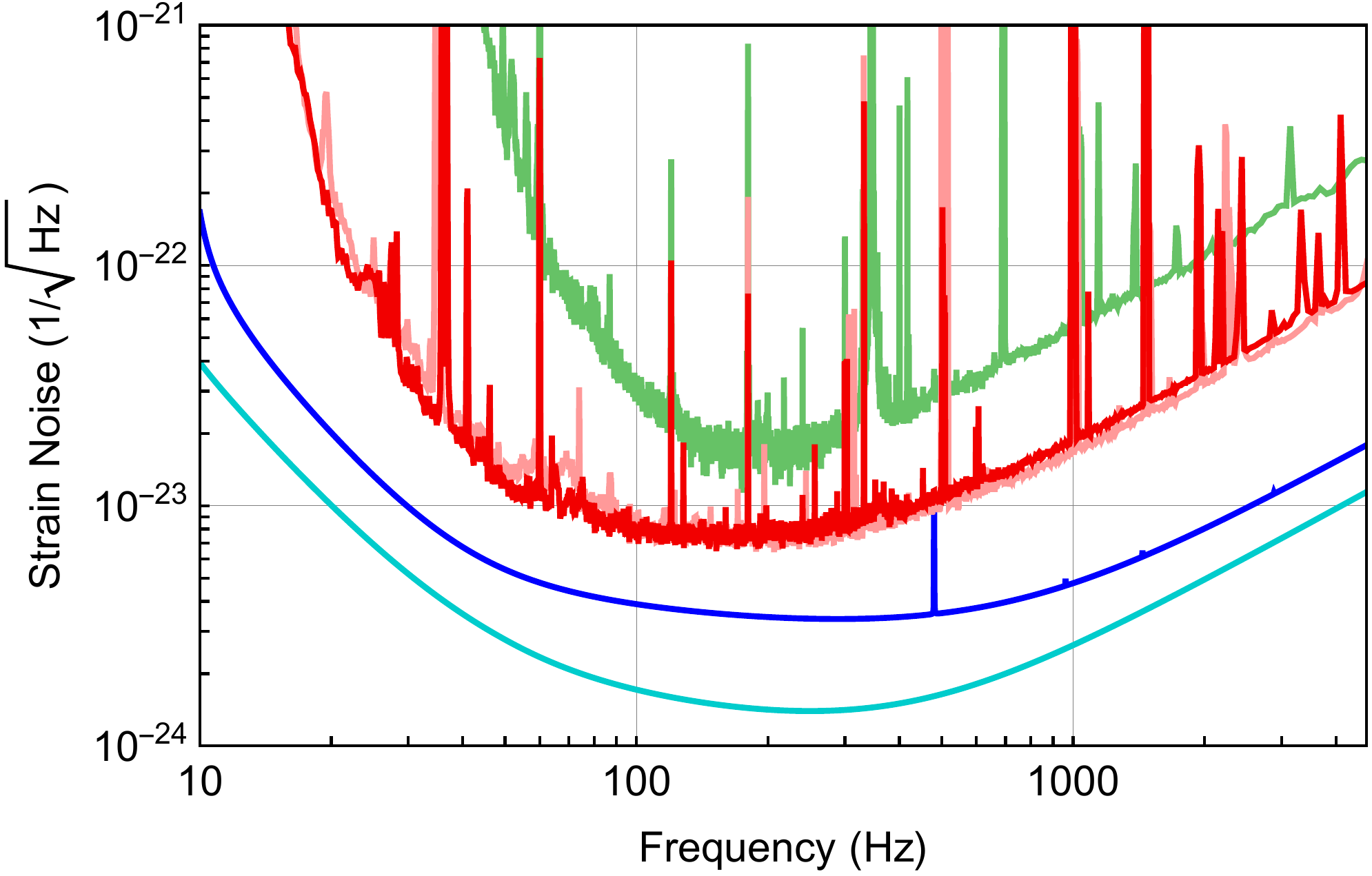}
    \end{minipage}
    \begin{minipage}{0.495\textwidth}
        \includegraphics[width = 0.95\textwidth]{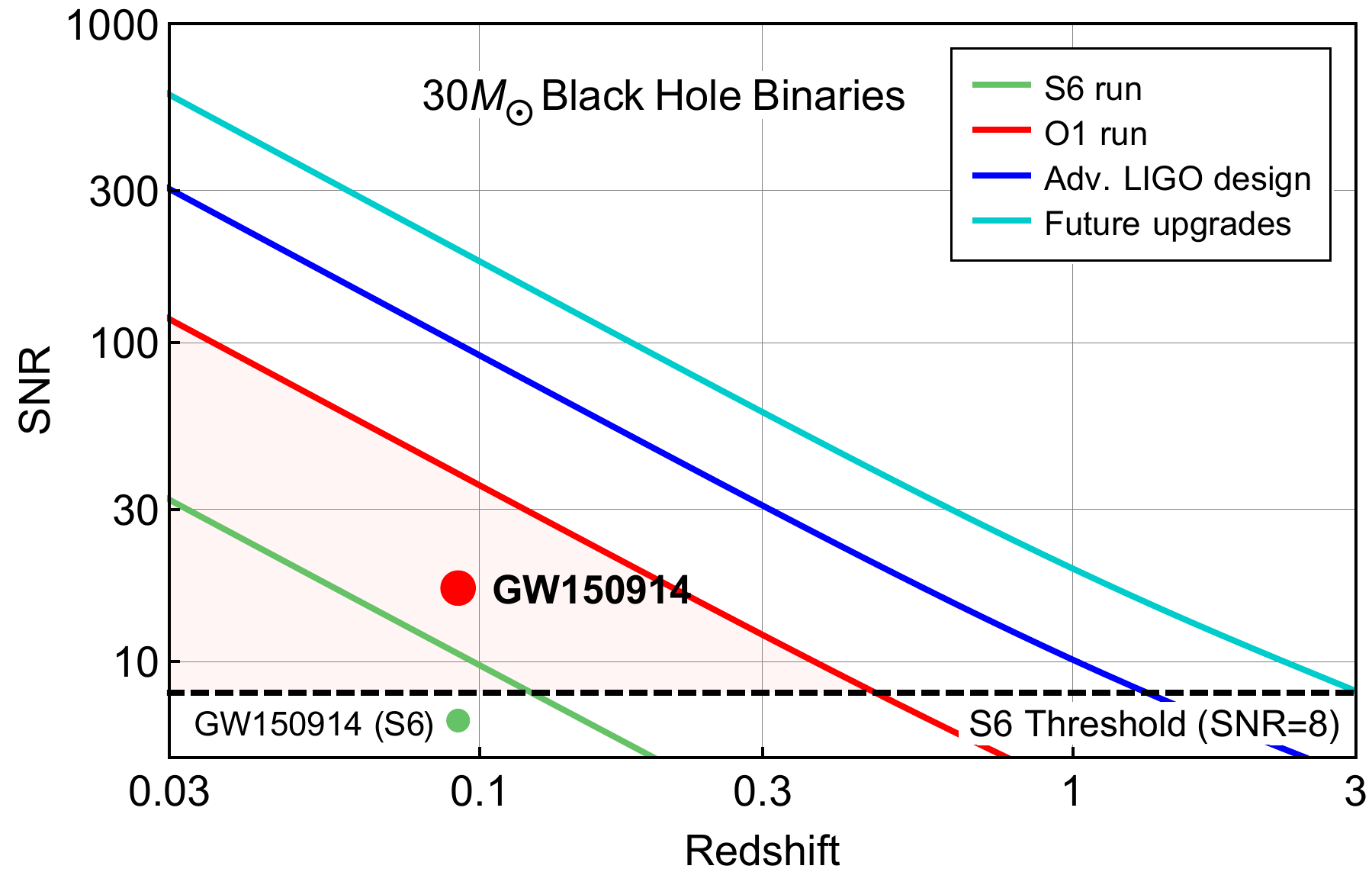}
    \end{minipage}
    \caption[Range]{The left plot shows the strain sensitivity during the first observation run (O1) of the Advanced LIGO detectors and during the last science run (S6) of the initial LIGO detectors. The O1 strain noise curve is shown for H1 (dark red) and L1 (light red); the two detectors have similar performance. The Advanced LIGO design sensitivity as well as a possible future upgrade~\cite{miller:2015} are shown to highlight the discovery potential in the coming years. The right plot shows the single detector signal-to-noise ratio (SNR) under optimal orientation as function of redshift $z$---for two merging black holes with mass 30\,$M_\odot$ each. GW150914 was not optimally orientated and was detected with a single detector SNR of 13 to 20 at $z=0.09$; this event would not have been seen in S6.}
    \label{fig:range}
\end{center}
\end{figure*}

{\it Astrophysical Reach} --- In general relativity, a gravitational wave far from the source can be approximated as a time-dependent perturbation of the space-time metric, expressed as a pair of dimensionless strain polarizations, $h_+$ and $h_\times$~\cite{thorne:1987}. An interferometric gravitational-wave detector acts as a transducer to convert these space-time perturbations into a measurable signal~\cite{weiss:1972}. The interferometer mirrors act as `freely falling' test masses. Advanced LIGO measures linear differential displacement along the arms which is proportional
to the gravitational-wave strain amplitude. We define the differential displacement as $\Delta L = \delta L_x - \delta L_y$, where $L_x = L_y = L$ are the lengths of two orthogonal interferometer arms. The gravitational-wave strain and the interferometer displacement are related through the simple equation $\Delta L = h\,L$, where $h$ is a linear combination of $h_+$ and $h_{\times}$.

The tiny displacements induced by astrophysical events demand that the interferometer mirrors be free from environmental disturbances and require a highly sensitive interferometric transducer---designed to be limited only by disturbances arising from fundamental physics consideration. Since the interferometer response to displacement, or equivalently gravitational-wave strain, is frequency dependent, it is necessary to represent the limiting detector noises as functions of frequency normalized by the interferometer response.

The left panel of Figure~\ref{fig:range} shows the amplitude spectral density of the total strain noise in units of strain per $\sqrt{\mathrm{Hz}}$ during the first observation run (O1 run) and, for comparison, during the final science run of the initial LIGO detectors (S6 run). In the detectors' most sensitive frequency band between 100\,Hz and 300\,Hz, the O1 strain noise is 3 to 4 times lower than achieved in the S6 run. At 50\,Hz, the improvement is nearly a factor of 100.

The right panel of Figure~\ref{fig:range} shows the single detector signal-to-noise ratio (SNR) for an optimally oriented compact binary system consisting of two 30\,$M_\odot$ black holes as a function of redshift $z$, and for different interferometer configurations. The observed strain amplitude is largest for a source whose orbital plane is parallel to the detector's plane and is located straight above or below; we refer to such a source as optimally oriented. The SNR is computed in the frequency domain~\cite{finn:1996} using standard cosmology~\cite{planck:2013} and phenomenological waveforms which account for inspiral, merger and ringdown, but not spins~\cite{ajith:2011}.

A Michelson interferometer lacks good directional sensitivity to gravitational waves. The antenna pattern covers approximately half the sky, both above and beneath the plane of the detector. Moreover, the antenna patterns of the two LIGO detectors are aligned to maximize the coincident detection of gravitational-wave signals, constrained to the 10\,ms inter-site propagation time. The coincidence constraint substantially rejects non-Gaussian noise and vetoes local transients. 

The observed strain amplitude is inversely proportional to the luminosity distance. For small redshifts, $z < 1$, the observable volume, and thus the detection rate, grows as the cube of the detector sensitivity. The number of detected events is expected to scale with the product of observing volume and observing time. Between September 12 and October 20 the H1 and L1 detectors had a duty cycle of 70\% and 55\%, respectively, while the observing time in coincidence was 48\%. After data quality processing~\cite{compDetChar:2016}, \fixme{16} days of data were analyzed around GW150914, resulting in a time-volume product of \fixme{0.1}\,$\rm{Gpc}^3$yr for binary black hole systems with masses similar to GW150914~\cite{compRate:2016}.

{\it The Displacement Measurement} --- The current generation of advanced detectors uses two pairs of test masses as coordinate reference points to precisely measure the distortion of the space-time between them. A pair of input and end test masses is located in each of the two arms of a Michelson laser interferometer, as shown in Figure~\ref{fig:setup}. The Advanced LIGO test masses are very pure and homogeneous fused silica mirrors of 34\,cm diameter, 20\,cm thickness and 40\,kg mass.

It is critical that the test masses be free from sources of displacement noise, such as environmental disturbances from seismic noise, or thermally driven motion. These noise sources are most relevant at frequencies below $100$\,Hz, while shot noise of the optical readout is dominant at high frequency. Figure~\ref{fig:noise} shows the measured displacement noise of Advanced LIGO during the first observing run, together with the major individual contributions, as discussed below.

To minimize ground vibrations, the test masses are suspended by multi-stage pendulums~\cite{aston:2012}, thus acting as free masses well above the pendulum resonance frequency of 0.4\,Hz. Monolithic fused silica fibers~\cite{braginski1992} are incorporated at the bottom stage to minimize suspension thermal noise~\cite{cumming:2012}, which limits the useful frequencies to 10\,Hz and above. The Advanced LIGO test masses require about 10 orders of magnitude suppression of ground motion above 10\,Hz. The multi-stage pendulum system attenuates the ground motion by seven orders of magnitude. It is mounted on an actively controlled seismic isolation platform which provides three orders of magnitude of isolation of its own~\cite{matichard:2015,wen:2014}. Moreover, these platforms are used to reduce the very large displacements produced by tidal motion and microseismic activity. Tidal forces can produce displacements up to several 100\,$\mu$m over a multi-kilometer baseline on time scales of hours. The dominant microseismic activity is driven by ocean waves. The resulting ground motion can be as large as several $\mu$m at frequencies around 0.15\,Hz---even far inland.

The entire test mass assembly including the suspension system and part of the seismic isolation system resides inside an ultra-high vacuum system, with pressures typically below 1\,$\mu$Pa over the $10,000\,{\rm m}^3$ volume, to prevent acoustic shorting of the seismic isolation systems and to minimize Rayleigh scattering in the optical readout.

The test masses are also susceptible to changes in the local gravitational field caused by changing mass distributions in their vicinity. While not limiting presently, at design sensitivity this time-dependent Newtonian noise source possibly becomes relevant below 20\,Hz, and might require active cancellation~\cite{cella:2000,driggers:2012}.

\begin{figure}[t]
\begin{center}
    \includegraphics[width = 0.485\textwidth]{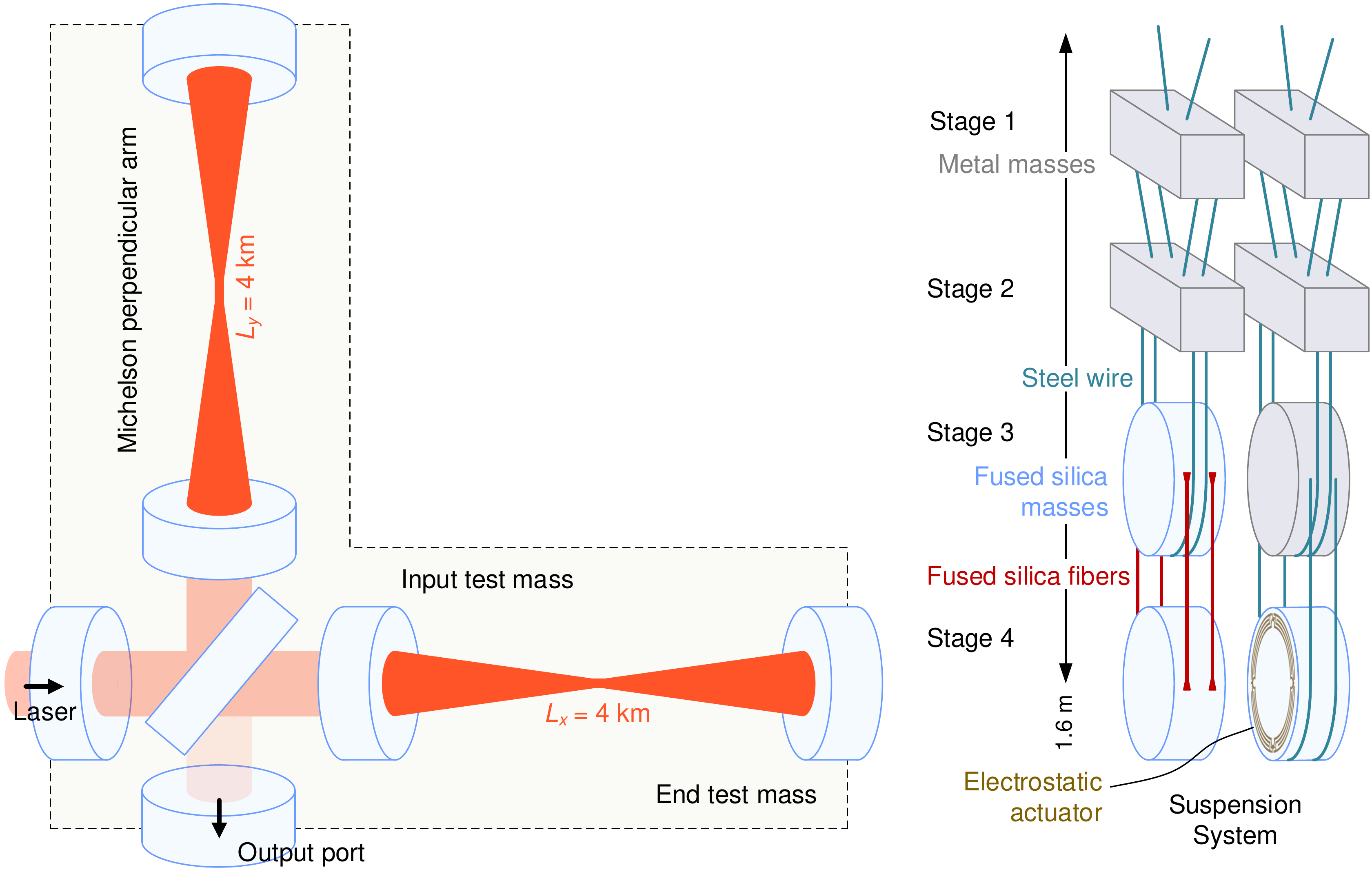}
    \caption[Setup]{Interferometer configuration and test mass setup. Each arm of the Michelson interferometer includes two suspended test masses. The two test masses are placed 4~km apart and form an optical resonator with a gain of 300. The suspension system is shown on the right, each test mass is at the bottom of a quadruple pendulum. It provides high isolation above the resonance frequencies which range from 0.4~Hz to 13~Hz. The test mass is attached to the penultimate mass through fused silica fibers providing a high mechanical quality factor which lowers the thermal noise. The other stages use steel wire. The attachment point to the seismic isolation system as well as stages~1 and~2 implement cantilever springs for vertical isolation. Each test mass is accompanied by its own reaction chain to minimize actuation noise. Coil actuators are mounted to the upper stages of the reaction chain and an electrostatic actuator is implemented at the bottom stage. Shown on the left are the other optics of the Michelson interferometer with the beamsplitter and the perpendicular arm. The two optics at the interferometer input and output port comprise the coupled resonator system which amplifies the response of the optical transducer.}
    \label{fig:setup}
\end{center}
\end{figure}

Thermally driven motion is another important source of displacement noise. It includes the Brownian motion of the suspension system~\cite{saulson:1990} as well as the test masses~\cite{levin:1998}, and mechanical loss in the mirror optical coatings~\cite{harry:2012}. The mirror coatings, a dielectric multilayer of silica and titania-doped tantala~\cite{harry:2007, granata:2016}, were developed to provide the required high reflectivity while minimizing coating thermal noise~\cite{flaminio:2010,agresti:2006,villar:2010}; it limits the design sensitivity in the central frequency band~\cite{aligo}.

The predicted levels for seismic, thermal and Newtonian noise sources are summarized in Figure~\ref{fig:noise} and compared to the total measured displacement noise. They are currently not limiting the sensitivity due to the presence of other technical noise sources, as detailed in Ref.~\cite{o1noise:2016}.

Quantum noise in the interferometer arises from the discrete nature of photons and their Poisson-distributed arrival rate~\cite{caves:1980, *caves:1981,Braginsky_1992,mcclelland:2011}. The momentum transfer of individual photons hitting a test mass gives rise to radiation pressure noise. Quantum radiation pressure noise scales as $1/mf^2$, where $m$ is the mass of the mirror and $f$ the frequency, and therefore it is most significant at lower frequencies.

Photon shot noise arises from statistical fluctuations in the photon arrival time at the interferometer output, and it is a fundamental limit of the transducer in sensing changes of the differential arm length. The importance of shot noise decreases as the inverse square-root of the laser power circulating in the interferometer arms. During the first observing run, Advanced LIGO was operating with 100\,kW of circulating laser power. The corresponding quantum noise curve, comprising both low frequency radiation pressure noise and high frequency shot noise, is shown in Figure~\ref{fig:noise}; it is limiting at frequencies above 100\,Hz. In the upcoming years, we plan to increase the circulating laser power up to 750\,kW, and thus reducing the shot noise contribution.

Coincident detection between the two LIGO observatories is used to reject transient environmental disturbances. Both observatory sites deploy seismometers, accelerometers, microphones, magnetometers, radio receivers, weather sensors, AC-power line monitors, and a cosmic ray detector for vetoes and characterization of couplings~\cite{effler:2015}.

{\it Interferometric Transducer} --- The Advanced LIGO detector uses a modified Michelson laser interferometer to translate strain into an optical phase shift~\cite{aligo}. Similar to an electromagnetic receiver, the optimal antenna length for a gravitational-wave detector is a quarter wavelength. For a gravitational wave at 100\,Hz this is 750\,km. The Advanced LIGO interferometer arms are 4\,km long and employ an optical resonator between the input and end test masses that multiplies the physical length by the effective number of round-trips of the light. However, the physical length cannot be arbitrarily short, because test mass displacement noises are multiplied by the same factor.

The output port of the Michelson interferometer is held at an offset from a dark fringe, resulting in a small amount of light leaving the output port~\cite{fricke:2012}. A differential optical phase shift will then decrease or increase the amount of light, depending which interferometer arm is momentarily stretched or squeezed by a passing gravitational wave. This light signal is measured by a photodetector, digitized and calibrated~\cite{compCalib:2016}, before being sent to the analysis pipelines~\cite{compBurst:2016,compCBC:2016}.

The calibration factor that converts detected laser light power to mirror displacement is measured by applying a known force to a test mass~\cite{goetz:2009}. An auxiliary 1047-nm wavelength laser is reflected off the end test mass and modulated in intensity to generate a varying force. The response of the optical transducer is measured by sweeping the modulation frequency through the entire detection band. It is also tracked by a set of fixed frequency lines. This way, the calibrated strain readout is computed in real-time with less than 10\% uncertainty in amplitude. The overall variability of the detector's sensitivity was about $\pm 10\%$.

\begin{figure}[t]
\begin{center}
    \includegraphics[width = 0.475\textwidth]{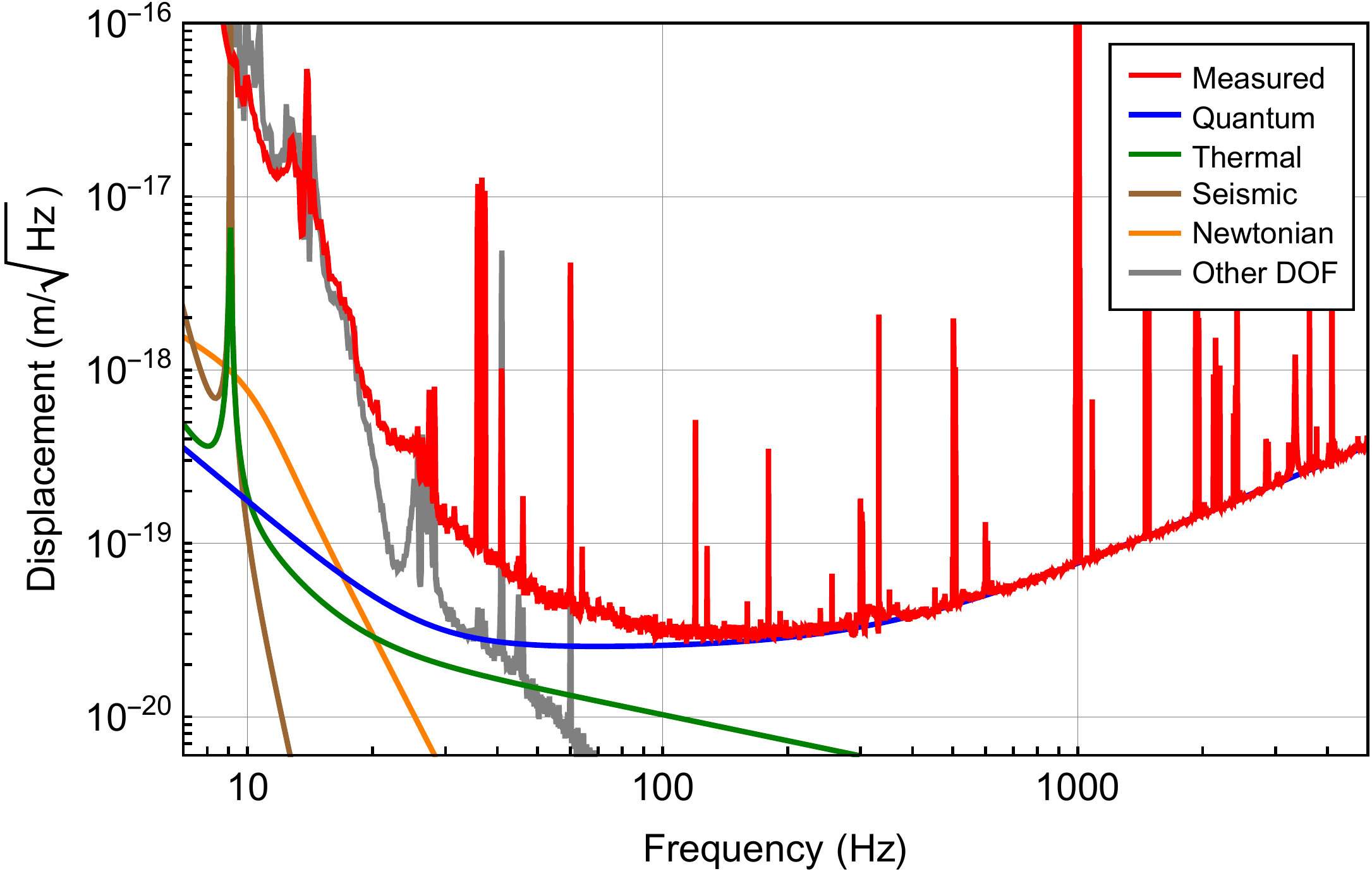}
    \caption[NoiseBudget]{The displacement sensitivity of the Advanced LIGO detector in Hanford during the first observation run O1; the Livingston detector has a similar sensitivity, as shown in Figure~\ref{fig:range}. The sum of all known noise sources accounts for most of the observed noise with the exception of the frequency band between 20\,Hz and 100\,Hz. This will be the focus of future commissioning to full sensitivity. The quantum noise includes both shot noise and radiation pressure noise. Thermal noise includes terms due to the suspensions, the test masses and the coatings. Seismic noise is the ground displacement attenuated through the seismic isolation system and the suspensions. Cross couplings from the auto-alignment system and from the auxiliary lengths are combined into the trace labelled ``other DOF'' (degrees-of-freedom). Newtonian gravitational noise is estimated from density perturbations due to surface ground motion. The strong line features are due to the violin modes of the suspension wires, the roll and bounce modes of the suspensions, the AC power line and its harmonics, and the calibration lines. Not shown are numerous noise sources that don't contribute significantly---such as laser frequency, intensity and beam jitter noise, sensor and actuation noise, and Rayleigh scattering by the residual gas~\cite{o1noise:2016}.}
    \label{fig:noise}
\end{center}
\end{figure}

The main light source is a pre-stabilized 1064-nm wavelength Nd:YAG laser. It is followed by a high power amplifier stage, capable of generating a maximum output power of 180\,W~\cite{Kwee:12}. During the first observation run, only 20\,W were injected into the interferometer. A triangular optical resonator of 32.9\,m round-trip length is placed between the laser source and the interferometer to reject higher order transverse optical modes and to stabilize the laser frequency further~\cite{mueller:2015}. At the output port, a bow-tie optical resonator of 1.3\,m round-trip length is used to reject unwanted frequency components on the light. Optical curvature mismatch of the interferometer mirrors is caused by manufacturing imperfections and by thermal lensing due to heating from the main laser beam. A thermal compensation system provides active correction by means of ring heaters arranged around the test masses and a set of CO$_2$ lasers for central heating~\cite{brooks:09}.

The Advanced LIGO detector uses coupled optical resonators to maximize the sensitivity of the interferometric transducer. These optical resonators enhance the light power circulating in each arm while simultaneously optimizing the effective antenna length and the gravitational-wave signal bandwidth~\cite{drever:1991,drever:1983,shilling,meers:1988,mizuno:1993}. As the interferometer is held near a dark fringe, most of the light is reflected back to the laser source. Adding a partially transmissive mirror at the interferometer input forms an optical resonator, leading to a power gain of 35 to 40 at the beamsplitter. The optical resonator in the interferometer arms enhances the circulating power by another factor of 300. Thus, 20\,W of laser power entering the interferometer results in nearly 100\,kW circulating in each arm. A partially reflective mirror is also placed at the output port to enhance the signal extraction and to increase the detector bandwidth. The resulting differential pole frequency or detector bandwidth is $\simeq 335\,$Hz (H1) and $\simeq 390\,$Hz (L1)~\cite{o1noise:2016}.

All of these coupled optical resonators require servo controls to be brought and held on resonance~\cite{staley:2014}. The lengths of the optical resonators in the interferometer arms are stabilized to less than 100\,fm, whereas the lengths of the other coupled resonators are kept within 1~to 10\,pm~\cite{o1noise:2016}. Similarly, the interferometer test masses are aligned within tens of nanoradians relative to the optical axis for optimal performance. The noise arising from sensing and control of these extra degrees-of-freedom are combined together in the curve labeled ``other DOF'' in Figure~\ref{fig:noise}. The Pound-Drever-Hall reflection locking technique is used to sense the auxiliary longitudinal degrees-of-freedom~\cite{PDH:1983, fritschel:01}, while an interferometric wavefront sensing scheme is deployed for the alignment system~\cite{ward:94, barsotti:2010}. Digital servo systems are used to feed control signals back to actuators which steer the relative longitudinal positions and orientations of the interferometer mirrors. To prevent reintroducing ground motion onto the test masses, electrostatic actuators~\cite{affeldt:2014} are mounted to a second quadruple pendulum known as the reaction chain. Only test masses use reaction chains; all other interferometer mirrors use coil actuators mounted on a rigid structure surrounding the suspensions.

Servo controls are also necessary for damping the plethora of normal modes of the pendular suspensions and for stabilizing the seismic isolation system to an inertial reference frame. Moreover, at high laser power, optical springs introduce angular instabilities due to photon radiation pressure-induced torques acting on the mirrors~\cite{sidles:2006, dooley:13}, while the mirror acoustic modes introduce parametric instabilities~\cite{braginskiPI2002, evans:2015}. At the current laser power only one acoustic mode is unstable which can be tuned away by the ring heaters. Together with thermal heating, angular optical springs and multiple parametric instabilities are the main challenges that need to be overcome to increase the circulating laser power; both will require active damping for stable operations.

Overall, more than 300~digital control loops with bandwidths spanning from sub-Hz to hundreds of kHz are employed to keep each Advanced LIGO interferometer operating optimally during observation. The digital controls computers also serve as the data acquisition system that continuously writes on the order of $10^5$ channels of time series data to disk, at a rate of $\simeq\!12\,$MB/s. It is synchronized to GPS to better than 10\,$\mu$s~\cite{compCalib:2016}. A state-based automation controller provides hands-free running during operations.

{\it Outlook} --- The global gravitational-wave network will be significantly enhanced in the upcoming years. In 2016 Advanced LIGO will be joined by Advanced Virgo, the 3\,km detector located near Pisa, Italy~\cite{avirgo}. The Japanese KAGRA interferometer~\cite{kagra} and a possible third LIGO detector in India~\cite{indigo:2011} will provide a global network that allows for improved parameter estimation and sky localization~\cite{lrr:2016}. Achieving design sensitivity with the network of current detectors will define earthbound gravitational-wave astrophysics in the near future. Looking further ahead, we can envision current technologies leading to a factor of two improvement over the Advanced LIGO design sensitivity~\cite{miller:2015}, so that events such as GW150914 could be detected with SNRs up to \fixme{200}. More dramatic improvements will require significant technology development and new facilities.

\begin{acknowledgments}

{\it Acknowledgement} --- The authors gratefully acknowledge the support of the United States
National Science Foundation (NSF) for the construction and operation of the
LIGO Laboratory and Advanced LIGO as well as the Science and Technology Facilities Council (STFC) of the
United Kingdom, the Max-Planck-Society (MPS), and the State of
Niedersachsen/Germany for support of the construction of Advanced LIGO
and construction and operation of the GEO600 detector.
Additional support for Advanced LIGO was provided by the Australian Research Council.
The authors gratefully acknowledge the Italian Istituto Nazionale di Fisica Nucleare (INFN),
the French Centre National de la Recherche Scientifique (CNRS) and
the Foundation for Fundamental Research on Matter supported by the Netherlands Organisation for Scientific Research,
for the construction and operation of the Virgo detector
and the creation and support  of the EGO consortium.
The authors also gratefully acknowledge research support from these agencies as well as by
the Council of Scientific and Industrial Research of India,
Department of Science and Technology, India,
Science \& Engineering Research Board (SERB), India,
Ministry of Human Resource Development, India,
the Spanish Ministerio de Econom\'ia y Competitividad,
the Conselleria d'Economia i Competitivitat and Conselleria d'Educaci\'o, Cultura i Universitats of the Govern de les Illes Balears,
the National Science Centre of Poland,
the European Commission,
the Royal Society,
the Scottish Funding Council,
the Scottish Universities Physics Alliance,
the Hungarian Scientific Research Fund (OTKA),
the Lyon Institute of Origins (LIO),
the National Research Foundation of Korea,
Industry Canada and the Province of Ontario through the Ministry of Economic Development and Innovation,
the Natural Science and Engineering Research Council Canada,
Canadian Institute for Advanced Research,
the Brazilian Ministry of Science, Technology, and Innovation,
Russian Foundation for Basic Research,
the Leverhulme Trust,
the Research Corporation,
Ministry of Science and Technology (MOST), Taiwan
and
the Kavli Foundation.
The authors gratefully acknowledge the support of the NSF, STFC, MPS, INFN, CNRS and the
State of Niedersachsen/Germany for provision of computational resources.

This document has been assigned the LIGO Laboratory document number LIGO-P1500237.
\end{acknowledgments}

\bibliography{detector}

\iftoggle{endauthorlist}{
  \let\author\myauthor
  \let\affiliation\myaffiliation
  \let\maketitle\mymaketitle
  \onecolumngrid
  \title{Authors}
  \pacs{}
  
  \newpage
  \date[\relax]{}
  \maketitle
}

\end{document}